\documentstyle[epsf,prl,twocolumn,aps,graphicx]{revtex}
\begin{document}
\title{ Symmetry Dependence of  Localization  in  Quasi- 1- dimensional Disordered Wires }
\author{S.\ Kettemann} 
\address
  { 
  I. Institut f\"ur Theoretische Physik, Universit\"at Hamburg,\\
  Jungiusstra\ss e 9, 20355 Hamburg}
\maketitle
\begin{abstract}

 The crossover in energy level statistics
  of a quasi-1-dimensional 
 disordered wire as a function of its length $L$ is used, 
 in order to derive 
    its averaged  localization length,  
 without   magnetic field, in a magnetic field and 
for moderate spin orbit scattering strength. 
  An analytical function of the magnetic field 
 for the local level spacing 
is obtained, 
 and  found to be  in excellent agreement with 
 the magnetic field dependent activation energy, 
  recently measured  
 in low-mobility quasi-one-dimensional wires\cite{khavin}. 
  This formula can be used to extract  
 directly and accurately  the localization length 
  from magnetoresistance experiments.
  In general,  
  the local level spacing is shown to be proportional to the 
  excitation gap  of a virtual particle, moving on 
 a compact symmetric space.
Pacs- numbers:  72.15.Rn,73.20.Fz,02.20.Qs
\end{abstract}

In  disordered wires  
  quantum interference 
 results in localization of 
all states  for arbitrary disorder strength, if the  wire is 
infinite\cite{reviews,ef}.  
  It has been 
 discovered  that the localization length 
  depends on the global symmetry of the wire~\cite{larkin}:
 $L_c = \beta \pi \hbar \nu S D_0  $, where $\beta =1, 2, 4$,
 corresponding to no magnetic field, finite magnetic field, and 
  strong spin- orbit scattering  or  magnetic impurities,
 respectively. $ \nu (E) $ is the electronic 
 density of states in  the wire. 
 $ D_0 $ is the classical diffusion constant 
 of the electrons in the wire, and $S$ its crossection. 
  This result was obtained by calculating the spatial
 decay of the  density correlation function 
  for wires whose thickness exceeds the 
 mean free path $l$.
 Independently, it was obtained 
  by calculating the 
 transmission probability through thin, few channel  wires\cite{fok}.
 Recently, the doubling of the localization length was observed 
 in sub-micron thin wires of doped GaAs by Khavin, Gershenson and Bogdanov,
 who found a continously decreasing activation energy when the magnetic field 
 is increased, saturating indeed at one half of its field free value
\cite{khavin}. 
 However, the  crossover as function of 
 moderate symmetry breaking fields
 defied any attempt to study it with  nonperturbative methods.
   The only results  were based on  
  a heuristic approach by Bouchaud\cite{bouchaud},
 a kind of semiclassical analysis by Imry and Lerner\cite{lerner}
  and  numerical studies\cite{pichard,leadbeater}.
 Recently, however, Kolesnikov and Efetov\cite{kolesnikov} succeeded to 
tackle that complex problem and
derived the density- density correlation function with the
 supersymmetry method in the crossover regime.
  The result does  not scale with a single localization length.

 Here, the problem of the crossover is addressed
from a different perspective.
 The 
  statistics of  discrete energy levels
 of a finite coherent, disordered metal particle 
 is an efficient way  to characterize its properties 
~\cite{ef}.
 This can be studied by calculating a disorder averaged 
autocorrelation function   
between two energies at a distance
 $\omega$ in the energy level spectrum.
It is an oscillatory function whose amplitude decays
with a power law when the energy levels in the vicinity of 
 the central  energy $E$ are extended. A Gaussian 
 decaying function is a strong indication that all states 
are localized.
 The autocorrelation function of spectral 
determinants (ASD) 
is defined by   
$
C(\omega) = \bar{C}(\omega)/ \bar{C}(0),
\hspace{.5cm} 
\bar{C}(\omega) = \left<\mbox{det}( E + \omega/2 - H) \mbox{det}
 ( E- \omega/2 -H )\right>.
$ $E$ is  a central energy. 
   We consider the hamiltonian of disordered electrons
\begin{equation}
H= \left[{\bf p
}- q {\bf A} \right]^2/2m + V({\bf x}),
\end{equation} 
 where $q$ is the electron charge, $c$ the velocity of light.  
$V({\bf x})$ is taken to be a Gaussian distributed random function
$\left<V({\bf x})\right> = 0$, 
and 
$\left<V({\bf x}) V({\bf x'})\right> =(\Delta S L/2 \pi \tau) 
 \delta ({\bf x} - {\bf x'}),$
 which models randomly distributed, uncorrelated impurities in the sample.
 $\Delta$ is the mean level spacing,  
 $ 1/\tau$  the elastic scattering rate.  
 The vector potential is used in 
 the gauge ${\bf A} = (- B y, 0, 0 )$, where $x$ is the coordinate
 along  the wire of length L,
  $y$ the one  in the direction perpendicular both 
to the wire
 and the magnetic field ${\bf B}$,
 which is directed  perpendicular to the wire. 
 The angular brackets denote averaging over impurities.  

The ASD  appears as an intermediate step 
 in the 
 Grassman replica trick and as
  the compact sector of the   supersymmetric
 field theory of disordered systems 
 ~\cite{ef,larkin}. 
 It is  a non-self-averaging quantity.  
It has been  shown recently, that the ASD  can distinguish between 
 an uncorrelated  spectrum of localized states
 and a correlated spectrum of extended states in the vicinity of $E$.  
 This way, the  metal- insulator crossover as
 a function of the length of a  disordered wire
 has been explored with the ASD.  
 A localization length $\xi_c$  in  a moderate magnetic field 
 has been   derived as the crossover length scale
 and shown to coincide with $L_c(\beta =2 )$\cite{prb}.
 The ASD also yields qualitative information on the 
 location of  localized states in a quantum- Hall- system
\cite{prl}.
  
 The 
 impurity averaged 
 ASD can be written as a partition function ~\cite{prl}
\begin{equation} \label{part}
\bar{C} ( \omega ) = \mbox{Tr} \exp (
  - L  \bar{H} \left[ Q  \right]),
\end{equation}
  where $\bar{ H}$ is an  
 effective Hamiltonian of  matrices  Q
 on a compact manifold, determined by  
  the symmetries of the  Hamiltonian $H$ of disordered electrons.
 Thus, the problem  reduces to the one of finding the 
  spectrum of the effective Hamiltonian $\bar{H}$. 

 There is 
   a finite gap $E_G$ between  the ground state energy and the 
 energy of the next excited state of  $\bar{H} (\omega = 0) $.
  For a long wire, $ L E_G \gg 1$, the ASD becomes, according to  Eq.
 (\ref{part}),   
$C(\omega) = \exp ( - const. L \omega^2/E_G)$.
 This  is  typical for a  a spectrum of localized states\cite{prl}.
 In the other limit $L E_G \ll 1$, 
all modes of $\bar{H}$ do contribute to the trace in the partition function 
Eq. (\ref{part}) 
with equal weight, yielding  the correlation 
 function of a spectrum of extended states\cite{prb}.
Thus, the crossover  length  
$\xi_c \sim 1/E_G$  can be identified with an averaged localization length. 

  Being a product of two spectral determinants,  
 $4 \alpha$ -component Grassman fields are needed to 
 get the functional integral representation of the ASD.
 Here,  $\alpha=1$, when the Hamiltonian is independent of the spin 
 of the electrons, and each level is doubly spin degenerate. 
  There is one pair of Grassman fields for each determinant 
 in the ASD and each pair is composed of a Grassman field
 and its time reversed one, as obtained by complex conjugation.
 $\alpha=2$ has to be considered, when the Hamiltonian does depend
 on spin, as for the case with moderately strong magnetic impurity or
 spin- orbit scattering. 
 This necessitates the use of a vector of  a spinor and the corresponding
time reversed one.
 One notes  a  global  invariance
 under rotations of these vectors.

 Averaging over impurities according to Eq. (1),  
one gets an interacting theory of Grassman fields 
 with interaction strength $1/\tau$. It
can be 
decoupled with a Gaussian integral over $4 \alpha \times 4 \alpha$ 
 matrices  which preserve the global
 rotational invariance. 
   The Grassman fields can now be integrated out exactly. 
 Finding  
  the  saddle point of the  integral over matrices  and
 integrating out longitudinal (massive) modes, 
 for $\omega \ll 1/\tau$ and
 $1/\tau \gg \Delta$, 
 the ASD  reduces to  a functional integral over  transverse
 modes Q, being elements of respective symmetric spaces.
 For   disordered  wires whose thickness $W$  is larger 
 than  the mean free path   $ l $
 but  smaller than the magnetic length $W < l_B$,  
the action reduces to the one of  the one-dimensional 
compact  nonlinear sigma model  ~\cite{weg,elk},
\begin{equation}
\bar{C} ( \omega ) = \int \prod_{ x} \mbox{d} Q ({ x})  \exp ( - F[Q] ),
\end{equation}
\begin{eqnarray}\label{free}
F[Q] &=& \alpha  \frac{1 }{16} L_{C  U}
 \int_0^L \mbox{d}  x \mbox{Tr} \left[ ( { { \nabla_x}} Q({ x}))^2
- \frac{q^2}{\hbar^2}  \overline{y^2}
 B^2 [ Q, \tau_3]^2 \right] 
 \nonumber \\
&+& i \alpha \frac{\pi}{4}  \frac{\omega}{\Delta}
\int \frac{\mbox{d}  x}{ L} \mbox{Tr} \Lambda_3  Q({ x}).
\end{eqnarray}
where $L_{C U} = L_C (\beta = 2) = 2 \pi \hbar \nu S D_0$ is the
localization length in the wire in a moderately strong magnetic field 
\cite{larkin}.  
 The  integral over the 
 crossection $S$ of the wire is done, giving
    $ \overline{y^2}$. 
 Here, and in the following, 
$\Lambda_i$ are the  Pauli matrices in the subbasis
 of the left and the right spectral determinant,
$\tau_i$ the ones in the subbasis spanned by  time reversal 
and $\sigma_i$  the ones in the space spanned
 by the spinor, for  $i=1,2,3$.  
 The  fluctuations of the 
 matrices  $Q$ are transverse,  $Q^2 = 1$, and 
 restricted by global rotational invariance to
 $ Q^+ =Q$.
    For $\alpha=1$   in addition 
  $Q^T C = C Q$, where $ C = - i \tau_2$.
  This constrains   $Q$ to be an element of  the 
 group defined on the 
compact symplectic symmetric space Sp(2)$/$( Sp(1) $\times$ Sp(1)). 
 For a moderate magnetic field, $Q$  is reduced to a 
 $2 \times 2$- matrix by the   broken 
 time reversal symmetry.
  This reduces the space of Q to $ U(2)/(U(1) \times U(1) )$.
 For $\alpha=2$    the matrix $C$ is, due to the time reversal 
 of the spinor,  substituted 
 by $i \sigma_2 \tau_1$\cite{elk}.
 Both magnetic impurities and 
 spin-orbit scattering   reduce the Q matrix to unity in spin space.
 Thus, C has effectively the form $\tau_1$.
   The condition $Q^T C = C Q$ leads therefore to a new symmetry class, 
 when the spin symmetry is broken
 but the 
 time reversal symmetry remains intact. This is the case 
 for moderately strong spin-orbit scattering. Then, $Q$ are $4 \times 4$-
 matrices on the orthogonal 
 symmetric space $ O(4)/(O(2) \times O(2) )$
\cite{hikami}.
  With magnetic impurities both the spin and  time reversal
 symmetry is broken,  and the Q- matrices
 are in the unitary symmetric space $U(2)/(U(1) \times U(1) )$
 as  for a moderate magnetic field and spin degenerate
 levels. The difference in the prefactor 
 $\alpha$ in the action, Eq.(\ref{free}), remains.
 One can extend this approach to other compact
 symmetric spaces with physical realizations, 
 see Ref. \onlinecite{zirnbauer1} for a complete classification.
  For $\omega/\Delta  <  L_{C U}/L$,  the spatial variation of $Q$ 
 can be neglected 
  and one retains the same  ASD 
as  for random matrices 
 of orthogonal, unitary and  symplectic symmetry,
 characterizing the energy levels of an ergodic 
 particle without magnetic field, with magnetic field, 
 and with spin- orbit- scattering, respectively\cite{smi}.
 The confusing  fact,  that random orthogonal matrices
 result for the ASD 
in a functional integral over compact  symplectic Q- matrices and
 vice versa,    results from the sign change of the Grassman variables
 under time reversal.   

 We can  derive the corresponding Hamiltonian $\bar{H}$
 by means of the transfer matrix method, 
 reducing the one-dimensional integral in Eq. (3) 
  to a single functional integral. 
 Thus, the ASD is obtained in the 
 simple form of Eq. (\ref{part}),
  with  the effective Hamiltonian 
\begin{equation} \label{effha}
\bar{H}( \omega = 0) =  \frac{1}{\alpha L_{CU} } (- 4 \Delta_Q
- \frac{1}{16}  X^2 Tr_Q [Q,\tau_3]^2).
\end{equation}
  $\Delta_Q$ is that
 part of the Laplacian on the  
 symmetric space, which couples to $Tr[\Lambda_3 Q ]$.
 $X =  2 \pi \alpha L_{CU} (\overline{y^ 2})^{(1/2)} B/\phi_0$,
 where $\phi_0 = q/h$ is the flux quantum.

The problem is now  equivalent to a particle with ``mass''
   $ (\alpha/8) L_{CU} (E) $ moving on the symmetric space 
of $Q$ in a harmonic potential 
 with ``frequency'' $  X/(\alpha L_{C U})$,
and in an 
 external field $i \alpha (\pi/4) \omega/(L \Delta)$,  in ``time'' $x$,
 the coordinate along the wire.
To find the ASD as a function of $\omega$ and the length of the wire 
 $L$, one can do a Fourier analysis in terms of the spectrum and eigenfunctions
 of 
 the effective Hamiltonian at zero frequency, $\bar{H} ( \omega =0 )$
\cite{zirnbauer2}.
 For $ L \gg \xi_c$, only the gap between the ground  and 
 the first excited state of $\bar{H} ( \omega = 0 )$, determines 
 the ASD.

  Without magnetic field, $B= 0$, 
 the Laplacian  is 
\begin{eqnarray} \label{orth}
 \Delta_Q & = & \partial_{\lambda_C} 
 ( 1 - \lambda_C^2 ) \partial_{\lambda_C}
 + 2 \frac{1 - \lambda_C^2}{\lambda_C}   \partial_{\lambda_C}
\nonumber \\ 
& + & \frac{1}{\lambda_C^2}  \partial_{\lambda_D} 
 ( 1 - \lambda_D^2 ) \partial_{\lambda_D},
\end{eqnarray}
where $\lambda_{C,D} \in [-1,1]$.
 Its ground state is $1$ and the first excited state is $ \lambda_C
 \lambda_D$.
 Thus, the gap is 
\begin{equation}
E_G ( B = 0 ) = 16/L_{©C U}.
\end{equation}
 For moderate magnetic field, with the condition  
$ L_{C U} (\overline{y^2})^{(1/2)} B  \gg  \phi_0 = h/q$, all
  degrees of
 freedom  arising from time reversal invariance
 are frozen out,  due to the term  $Tr_Q [ Q, \tau_3]^2 = 16 (\lambda_C^2-1)$
 which fixes 
 $\lambda_C^2 = 1$. Then, the Laplacian reduces
 to 
\begin{eqnarray} \label{unit}
 \Delta_Q =  
\partial_{\lambda_D} 
 ( 1 - \lambda_D^2 ) \partial_{\lambda_D}.
\end{eqnarray}
 Its eigenfunctions are the Legendre polynomials. 
 There 
 is a  gap above the isotropic ground state
  of  magnitude
\begin{equation}
 E_G ( X \gg 1 ) =8/L_{C U}.
\end{equation}
   For moderate magnetic impurity scattering exceeding the local level spacing,
 $1/\tau_s > \Delta_C$, the Laplacian is 
 given by Eq.(8).
  Thus, due to $\alpha=2$, the gap is   reduced
 to $
E_G (1/\tau_S > \Delta_C) = 4/L_{C U}$.
 For moderately strong 
spin- orbit scattering $1 / \tau_{SO} > \Delta_C$,
 the Laplace operator is 
\begin{equation} 
\Delta_Q =\sum_{l=1,2} \partial_{\lambda_l} 
 ( 1 - \lambda_l^2 ) \partial_{\lambda_l},
\end{equation}
 where $\lambda_{1,2} \in [-1,1]$. 
 The ground state is $\psi_0 = 1$,  the first excited state
 is doubly degenerate, $\psi_{11} = \lambda_1$, 
 $\psi_{12} = \lambda_2$.
  Thus, the gap is the same as for magnetic impurities,
\begin{equation} 
 E_G ( 1/\tau_{SO} > \Delta_C ) = 4/L_{C U}.
\end{equation}
 An external  magnetic field lifts 
 this degeneracy but does not  change the gap.
  
  Using  the crossover in energy level statistics
 as the definition of a localization length 
 as  above, we get 
in a quasi- 1 -dim. wire,
\begin{equation} 
\xi_c = 1/E_G(\beta)= ( 1/16 ) \beta L_{C U} ,
\end{equation}
 where $\beta =1, 2, 4$
 corresponding to no magnetic field, finite magnetic field, and 
  strong spin- orbit scattering  or  magnetic impurities,
 respectively.  Comparing with the known equation for the 
localization length, $ L_c $, we find that 
 the dependence of the ratios  $\beta$ on the symmetry  are 
 in perfect 
agreement with the result as obtained from the 
 spatial decay of the density- density- correlation function\cite{larkin},
 while it defers by the overall constant $ 1/ 8$.
 
This relation can be proven directly. 
   The ASD at zero frequency  $\bar{C} ( 0 )_{L}$ of the wire of 
 length $L$, 
 becomes, when the wire is divided  into two parts,
 $\bar{C} ( 0 )_{L/2}^2$.
 For $L \rightarrow \infty$, we find that the relative difference
 is:
\begin{equation}\label{specdec}
f(L) = \frac{\bar{C} ( 0 )_{L/2}^2}{
\bar{C} ( 0 )_{L}}- 1  = 2 \exp ( - L E_G/2 ), 
\end{equation}
exponentially decaying with the length $L$. 
  $f(L)$ can be estimated, following an argument by Mott\cite{mott}: 
 When the two halves of the wire get connected, the Eigenstates of the 
 two separate halves become hybridized and the Eigenenergy 
 of a state $ \psi_n $ is changed
 by $ \pm \Delta_C \exp ( - 2 x_n/L_C ) $. 
 $x_n$ is random, 
 depending on the position of an eigenstate with closest energy in 
 the other half of the wire. 
  Thus, averaging over $x_n$  gives:
\begin{equation} 
f( L  ) \sim + \exp ( - 4 L/ L_C ).
\end{equation}   
 Comparison with Eq. (\ref{specdec}) yields indeed $ 1/L_C = 8 E_G$.
   
{ \it The Crossover Behaviour of the Localization Length}

  This direct relation of the ASD to the spatial 
 overlap between wavefunctions allows us to  extend this approach 
 to get an  analytical
 solution for   the 
 crossover behaviour of the localization 
 length and  the local level spacing  as a magnetic field is turned on, 
and without  strong spin- orbit scattering.
While a heuristic approach\cite{bouchaud} and
  numerical studies\cite{leadbeater} seemed to indicate 
 a continous increase of the localization length, the analytical result
\cite{kolesnikov} does show
 that both limiting localization lengths
 $L_c(\beta = 1)$ and $L_c(\beta =2)$ are present in the crossover regime
 and that there is no single parameter scaling.
 This is explained by  arguing that the far tails 
 of the wavefunctions do cover a large enough area to have fully broken 
 time reversal symmetry, decaying with the length scale $L_c (\beta =2)$
 even if the magnetic field 
 is too weak to affect the properties of 
 the  bulk of the 
 wavefunction, which  does decay at smaller length scales with the shorter
  localization length $L_c (\beta =1)$, corresponding to the 
 time reversal symmetric case.
The quantity studied there is the imprurity averaged correlation function 
 of local wavefunction amplitudes and its momenta 
 at a fixed energy $\epsilon$:
$Y(\epsilon) = < \sum_{\alpha} \mid \psi_{\alpha} (0) \mid^2 
\mid \psi_{\alpha} ( r) \mid^2 \delta (\epsilon - \epsilon_{\alpha} )>$.
 It is   averaged  over  a distribution 
 of eigenfunctions in different impurity representations. Thus,  
  each eigenfunction could decay exponentially 
 with a single localization length, but with a  distribution 
 which has two maxima, at  $L_c(\beta=1)$ and $L_c(\beta=2)$, 
 whose weight is a function of the magnetic field in the crossover regime.
  It is  interesting to ask if this property of the 
 eigenfunctions does have a  consequence on the energy level statistics 
 as well, or, if that is still governed by a single parameter as the magnetic 
field  is varied. 
 
 The effective Hamiltonian 
 for  moderate magnetic fields is given, without spin dependent scattering,
 $\alpha=1$, by:
\begin{equation}
\bar{H} =\frac{1}{L_{C U}} ( - 4 \Delta_Q + X^2 ( 1-\lambda_C^2) )  ,
\end{equation}
 where the Laplacian is 
 Eq.[\ref{orth}] and 
  $X =  2 \pi \phi/\phi_0 $, 
 where $\phi$ is the magnetic flux 
through an area $A =  L_{C U} (\overline{y^2})^(1/2)$.

In the limit $ X \rightarrow 0$ the ground state and first 
 excited state  approach $ 1, \lambda_C \lambda_D$, respectively.
In the limit $ X \gg 1$, $\lambda_C^2$ becomes fixed to 1.
 Thus, the 
Ansatz $\psi_0(\lambda_C ) = \exp ( A_0 (1- \lambda_C^2))$,
 and $\psi_1(\lambda_C,\lambda_D ) = \lambda_C \lambda_D \exp ( A_1 (1-\lambda_C^2))$,
 solves $\bar{H} \psi =\bar{ E} \psi$
 to first order in  $z= X^2 ( 1-\lambda_C^2)$ and yields the gap:
\begin{equation} \label{smooth}
E_G (X) = 4( 2 + \sqrt{49 +  X^2} - \sqrt{25 +  X^2})/L_{C U}.
\end{equation}
 This  solution is valid in both 
 the limits $X \ll 1$ and $X \gg 1$, interpolating 
  the region $ X \approx 1$.
 The resulting ratio of local energy level spacings 
 $\Delta_C(B)/\Delta_C(0) = E_G(B)/E_G(0)$,
 is  shown in Fig. 5 to be in excellent qualitative agreement with 
 experimental data for   
 the magnetic field dependent 
activation energy, measured recently in  transport experiments\cite{khavin}.

 There is a quantitative discrepancy by a factor $.6$
 between the best fit  $X = .036 H/Oe$,  
 and 
  $X =  2 \pi \phi/\phi_0 $, 
  $\phi = \mu_0 H L_{C U} (\overline{y^2})^{(1/2)}$.
 With the experimental 
 parameters   $ \alpha=1, L_{CO} = .61 \mu m$, width $W = .2 \mu m $ 
of  sample 5 in Ref. \onlinecite{khavin} 
 and   $ \overline{y^2}  = W^2/12$ for a 2- dimensional wire, it yields
 $ X= .021 H/Oe$. We note that smooth confinement can give  
 $\overline{y^2}  > W^2/12$,  thus explaining  this discrepancy
 and the observed difference between $W$ as obtained 
 from the sample resistance
 and estimated from the analysis of the weak localization magnetoresistance,
 which also depends on $\overline{y^2}$\cite{khavin2}.
  When $  W^2 B > \phi_0$ or $ H > H^* = 2.1 * 10^{-11} Oe [ m^2 /W^2]
 = 525 Oe$,
the derivation has to be extended to the 2- dimensional nonlinear sigma- model, 
 and Eq. \ref{smooth} seizes to be valid. Then,  $L_C$
  increases as function of magnetic field more strongly.

 The
 asymptotic behaviour, of $\delta L_C(B) \sim B^2$ for small  
 and $\sim 1/B$ at large magnetic fields of Eq. (\ref{smooth}) does agree with the suggestion
 by Bouchaud\cite{bouchaud}.
\begin{figure} \label{fig1}
\includegraphics[width=0.44 \textwidth]{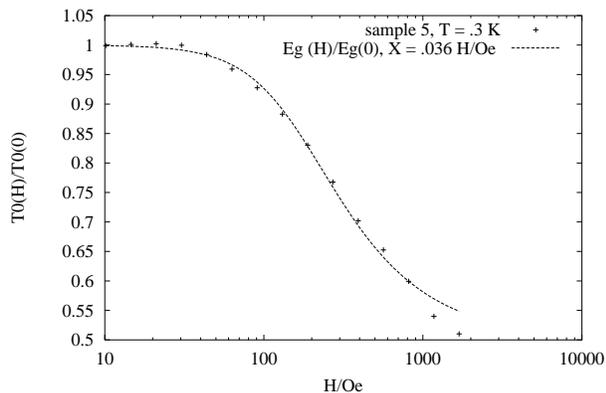}
\begin{center}
\caption{  The  activation gap ratio
  $T_0(H)/T_0(0)$ as a function of the magnetic field H in Oe
of sample 5 measured at temperature $T= .3 K$ as reported in Ref. 1, 
 together with the theoretical curve.     }
\end{center}
\end{figure}
  The two length scale physics of Ref. \onlinecite{kolesnikov}
 has thus 
 no consequence for the energy level statistics as studied with the 
 ASD. Since we could also show a direct relation between this spectral 
 statistics and the exponential decay of localized eigenfunctions, 
 it is  suggestive to be the non self averaging property 
 of the ASD which washes out the two scale physics.

The author gratefully acknowledges, 
 usefull comments by Isa Zarekeshev, 
 discussions with  Bernhard Kramer and  
  support from DFG.
  He thanks 
 Yuri Khavin for usefull communications and 
Martin Zirnbauer for pointing out a 
   serious confusion in  the  notation.

\end{document}